% ... May 4 by JAM
% ... May 2 by ES
%Draft edited April 27, 2007 by JAM
%Draft edited April 14, 2007 by AEL
\input harvmac

\noblackbox

\let\includefigures=\iftrue
\let\useblackboard=\iftrue
\newfam\black

%Figure Stuff
\includefigures
\message{If you do not have epsf.tex (to include figures),}
\message{change the option at the top of the tex file.}
\input epsf
\def\figin{\epsfcheck\figin}\def\figins{\epsfcheck\figins}
\def\epsfcheck{\ifx\epsfbox\UnDeFiNeD
\message{(NO epsf.tex, FIGURES WILL BE IGNORED)}
\gdef\figin##1{\vskip2in}\gdef\figins##1{\hskip.5in}% blank space instead
\else\message{(FIGURES WILL BE INCLUDED)}%
\gdef\figin##1{##1}\gdef\figins##1{##1}\fi}
\def\DefWarn#1{}
\def\figinsert{\goodbreak\midinsert}
\def\ifig#1#2#3{\DefWarn#1\xdef#1{Fig.~\the\figno}
\writedef{#1\leftbracket Fig.\noexpand~\the\figno}%
\figinsert\figin{\centerline{#3}}\medskip\centerline{\vbox{
\baselineskip12pt\advance\hsize by -1truein
\noindent\footnotefont{\bf Fig.~\the\figno:} #2}}
\bigskip\endinsert\global\advance\figno by1}
%%%
\else
\def\ifig#1#2#3{\xdef#1{Fig.~\the\figno}
\writedef{#1\leftbracket Fig.\noexpand~\the\figno}%
%\figinsert\figin{\centerline{#3}}\medskip
%\centerline{\vbox{\baselineskip12pt
%\advance\hsize by -1truein\noindent
%\footnotefont{\bf Fig.~\the\figno:} #2}}
%\bigskip\endinsert
\global\advance\figno by1} \fi

\def\doublefig#1#2#3#4{\DefWarn#1\xdef#1{Fig.~\the\figno}
\writedef{#1\leftbracket Fig.\noexpand~\the\figno}%
\figinsert\figin{\centerline{#3\hskip1.0cm#4}}\medskip\centerline{\vbox{
\baselineskip12pt\advance\hsize by -1truein
\noindent\footnotefont{\bf Fig.~\the\figno:} #2}}
\bigskip\endinsert\global\advance\figno by1}

%%BLACKBOARD FONT STUFF
\useblackboard
\message{If you do not have msbm (blackboard bold) fonts,}
\message{change the option at the top of the tex file.}
\font\blackboard=msbm10 scaled \magstep1 \font\blackboards=msbm7
\font\blackboardss=msbm5 \textfont\black=\blackboard
\scriptfont\black=\blackboards \scriptscriptfont\black=\blackboardss

\else

\fi
% *************************************
%\draft
%
\def\subsubsec#1{\bigskip\noindent{\it{#1}} \bigskip}
\def\yboxit#1#2{\vbox{\hrule height #1 \hbox{\vrule width #1
\vbox{#2}\vrule width #1 }\hrule height #1 }}
\def\fillbox#1{\hbox to #1{\vbox to #1{\vfil}\hfil}}
\def\ybox{{\lower 1.3pt \yboxit{0.4pt}{\fillbox{8pt}}\hskip-0.2pt}}
%
%
%%MATH MACROS
%Greek letters and their bars

%manifold and its universal cover

\def\M{{\cal M}}

\def\J{{\cal J}}

%More bars

%\def\l{\left}
%\def\r{\right}
\def\comments#1{}

\def\p{\partial}

\def\half{{1\over 2}}

\def\bra#1{{\langle}#1|}
\def\ket#1{|#1\rangle}
\def\brket#1#2{\langle #1 | #2 \rangle}

\def\CM{{\cal M}}
\def\CN{{\cal N}}
\def\CO{{\cal O}}%AEL

\def\CW{{\cal W}}
%AEL

\def\II{\relax{I\kern-.10em I}}

\def\IZ{\relax{\rm Z\kern-.34em Z}}
\def\IB{\relax{\rm I\kern-.18em B}}
\def\IC{{\relax\hbox{$\inbar\kern-.3em{\rm C}$}}}
\def\ID{\relax{\rm I\kern-.18em D}}
\def\IE{\relax{\rm I\kern-.18em E}}
\def\IF{\relax{\rm I\kern-.18em F}}
\def\IG{\relax\hbox{$\inbar\kern-.3em{\rm G}$}}
\def\IGa{\relax\hbox{${\rm I}\kern-.18em\Gamma$}}
\def\IH{\relax{\rm I\kern-.18em H}}
\def\II{\relax{\rm I\kern-.18em I}}
\def\IK{\relax{\rm I\kern-.18em K}}
\def\IP{\relax{\rm I\kern-.18em P}}
%\def\IX{\relax{\rm X\kern-.01em X}}
%this doesn't work

%

\def\inbar{\,\vrule height1.5ex width.4pt depth0pt}

\def\p{\partial}

\def\IR{\relax{\rm I\kern-.18em R}}

\def\simgt{\hskip0.05in\relax{
\raise3.0pt\hbox{ $>$ {\lower5.0pt\hbox{\kern-1.05em $\sim$}} }}
\hskip0.05in}

%

 % for now

%

\def\lp10{\ell_p^{10}}
\def\lp11{\ell_p^{11}}
\def\R11{R_{11}}

\def\frac#1#2{{#1 \over #2}}

%identity operator from doyon-fonseca

%% from the topological vertex paper

%%                              TABLEAUX.TEX
%%      This  macro file is for producing a ``Young Tableau'' which is
%%      an array of little squares sometimes used in mathematical physics.
%%      For instance, the command $\tableau{6 3 2}$ will produce a tableau
%%      with 6 squares in the top row, 3 in the next, and 2 in the last.
%%                                  OOOOOO
%%      This tableau will look like OOO    but made of squares instead of O's.
%%                                  OO
%%      Any number of rows may be present, each having a nonzero number of
%%      squares.
%%
%%      A tableau is math mode material, so use $ or $$ to enclose it.
%%
%%      The size and line-thickness of the little boxes are controlled by the
%%      dimension parameters --
%%              \tableauside=1.0ex              %(size)
%%              \tableaurule=0.4pt              %(line-thickness)
%%      Change them if you want.
%%
%%                                                      -- Doug Eardley 9/19/8%%
%%
\newdimen\tableauside\tableauside=1.0ex
\newdimen\tableaurule\tableaurule=0.4pt
\newdimen\tableaustep
\def\phantomhrule#1{\hbox{\vbox to0pt{\hrule height\tableaurule width#1\vss}}}
\def\phantomvrule#1{\vbox{\hbox to0pt{\vrule width\tableaurule height#1\hss}}}
\def\sqr{\vbox{%
  \phantomhrule\tableaustep
  \hbox{\phantomvrule\tableaustep\kern\tableaustep\phantomvrule\tableaustep}%
  \hbox{\vbox{\phantomhrule\tableauside}\kern-\tableaurule}}}
\def\squares#1{\hbox{\count0=#1\noindent\loop\sqr
  \advance\count0 by-1 \ifnum\count0>0\repeat}}
\def\tableau#1{\vcenter{\offinterlineskip
  \tableaustep=\tableauside\advance\tableaustep by-\tableaurule
  \kern\normallineskip\hbox
    {\kern\normallineskip\vbox
      {\gettableau#1 0 }%
     \kern\normallineskip\kern\tableaurule}%
  \kern\normallineskip\kern\tableaurule}}
\def\gettableau#1 {\ifnum#1=0\let\next=\null\else
  \squares{#1}\let\next=\gettableau\fi\next}

\tableauside=1.0ex \tableaurule=0.4pt

%% from shiraz

 %
 %       \eqn\label{a+b=c}       gives displayed equation, numbered
 %                               consecutively within sections.
%     \eqnn and \eqna define labels in advance (of eqalign?)
 %
 \def\eqnn#1{\xdef #1{(\secsym\the\meqno)}\writedef{#1\leftbracket#1}%
 \global\advance\meqno by1\wrlabeL#1}
 \def\eqna#1{\xdef #1##1{\hbox{$(\secsym\the\meqno##1)$}}
 \writedef{#1\numbersign1\leftbracket#1{\numbersign1}}%
 \global\advance\meqno by1\wrlabeL{#1$\{\}$}}
 \def\eqn#1#2{\xdef #1{(\secsym\the\meqno)}\writedef{#1\leftbracket#1}%
 \global\advance\meqno by1$$#2\eqno#1\eqlabeL#1$$}

\global\newcount\itemno \global\itemno=0

%\bigbreak}
%\bigskip\noindent}
\def\itemaut#1{\global\advance\itemno by1\noindent\item{\the\itemno.}#1}

%\itemized
%\itemaut{First this.}
%\itemaut{Then that.}

%%ENGLISH MACROS
\def\eg{{\it e.g.}}

\hyphenation{Di-men-sion-al}

%%REFERENCING MACROS

%%

\lref\MPtowards{
  D.~R.~Morrison and M.~R.~Plesser,
  ``Towards mirror symmetry as duality for two dimensional abelian gauge
  theories,''
  Nucl.\ Phys.\ Proc.\ Suppl.\  {\bf 46}, 177 (1996)
  [arXiv:hep-th/9508107].
  %%CITATION = NUPHZ,46,177;%%
}

\lref\sing{ L.~Fidkowski, V.~Hubeny, M.~Kleban and S.~Shenker,
  ``The black hole singularity in AdS/CFT,''
  JHEP {\bf 0402}, 014 (2004)
  [arXiv:hep-th/0306170];
  %%CITATION = JHEPA,0407,073;%%
  T.~Hertog and G.~T.~Horowitz,
  ``Towards a big crunch dual,''
  JHEP {\bf 0407}, 073 (2004)
  [arXiv:hep-th/0406134].
%%CITATION = JHEPA,0402,014;%%
S.~R.~Das, J.~Michelson, K.~Narayan and S.~P.~Trivedi,
  ``Cosmologies with null singularities and their gauge theory duals,''
  Phys.\ Rev.\  D {\bf 75}, 026002 (2007)
  [arXiv:hep-th/0610053].
  %%CITATION = PHRVA,D75,026002;%%
}

%\CrapsWD
\lref\CrapsWD{
  B.~Craps, S.~Sethi and E.~P.~Verlinde,
  ``A matrix big bang,''
  JHEP {\bf 0510}, 005 (2005)
  [arXiv:hep-th/0506180];
  %%CITATION = JHEPA,0510,005;%%
T.~Ishino, H.~Kodama and N.~Ohta,
  ``Time-dependent solutions with null Killing spinor in M-theory and
  superstrings,''
  Phys.\ Lett.\  B {\bf 631}, 68 (2005)
  [arXiv:hep-th/0509173].
  %%CITATION = PHLTA,B631,68;%%
}

\lref\farkaskra{ H.~M.~Farkas and I.~Kra,
{\it Riemann surfaces}, Graduate Texts in Mathematics, vol. 71,
Springer-Verlag, New York, 1980.}

\lref\kawauchi{ A.~Kawauchi,
``Mutative hyperbolic homology {$3$}-spheres with the same {F}loer homology,''
Geom. Dedicata {\bf 61}, 205 (1996).
}

\lref\masters{ J. D. Masters,
``Virtual {B}etti numbers of genus 2 bundles,''
Geom. Topol. {\bf 6}, 541 (2002) [arXiv:math.GT/0201118].
}

\lref\susycon{ S.~J.~Gates, M.~T.~Grisaru, M.~Rocek and W.~Siegel,
  ``Superspace, or one thousand and one lessons in supersymmetry,''
  Front.\ Phys.\  {\bf 58}, 1 (1983)
  [arXiv:hep-th/0108200].
  %%CITATION = FRPHA,58,1;%%
(the dimensional reduction of the 3d $N=1$ ``toy superspace" of chapter 2)}

\lref\TFA{ A.~Adams, X.~Liu, J.~McGreevy, A.~Saltman and E.~Silverstein,
  ``Things fall apart: Topology change from winding tachyons,''
  JHEP {\bf 0510}, 033 (2005)
  [arXiv:hep-th/0502021].
  %%CITATION = JHEPA,0510,033;%%
}

\lref\Coleman{
%\ColemanUZ
  S.~R.~Coleman,
  ``More About The Massive Schwinger Model,''
  Annals Phys.\  {\bf 101}, 239 (1976).
  %%CITATION = APNYA,101,239;%%
}

\lref\HV{ K.~Hori and C.~Vafa,
  ``Mirror symmetry,''
  arXiv:hep-th/0002222.
  %%CITATION = HEP-TH/0002222;%%
}

\lref\GMV{ R.~Gregory, J.~A.~Harvey and G.~W.~Moore,
  ``Unwinding strings and T-duality of Kaluza-Klein and H-monopoles,''
  Adv.\ Theor.\ Math.\ Phys.\  {\bf 1}, 283 (1997)
  [arXiv:hep-th/9708086].
  %%CITATION = 00203,1,283;%%
}

\lref\gary{ G.~T.~Horowitz and E.~Silverstein,
  ``The inside story: Quasilocal tachyons and black holes,''
  Phys.\ Rev.\  D {\bf 73}, 064016 (2006)
  [arXiv:hep-th/0601032].
  %%CITATION = PHRVA,D73,064016;%%
}

\lref\OV{ H.~Ooguri and C.~Vafa,
  ``Two-Dimensional Black Hole and Singularities of CY Manifolds,''
  Nucl.\ Phys.\ B {\bf 463}, 55 (1996)
  [arXiv:hep-th/9511164].
  %%CITATION = HEP-TH 9511164;%%
  }

%\KrausHV
\lref\KrausHV{
  P.~Kraus, F.~Larsen and S.~P.~Trivedi,
  ``The Coulomb branch of gauge theory from rotating branes,''
  JHEP {\bf 9903}, 003 (1999)
  [arXiv:hep-th/9811120].
  %%CITATION = HEP-TH 9811120;%%
}

%\DanielssonFA
\lref\DanielssonFA{
  U.~H.~Danielsson, E.~Keski-Vakkuri and M.~Kruczenski,
  ``Black hole formation in AdS and thermalization on the boundary,''
  JHEP {\bf 0002}, 039 (2000)
  [arXiv:hep-th/9912209].
  %%CITATION = HEP-TH 9912209;%%
}

%\GiddingsZU
\lref\GiddingsZU{
  S.~B.~Giddings and S.~F.~Ross,
  ``D3-brane shells to black branes on the Coulomb branch,''
  Phys.\ Rev.\ D {\bf 61}, 024036 (2000)
  [arXiv:hep-th/9907204].
  %%CITATION = HEP-TH 9907204;%%
  }

\lref\Edphases{ E.~Witten,
  ``Phases of N = 2 theories in two dimensions,''
  Nucl.\ Phys.\  B {\bf 403}, 159 (1993)
  [arXiv:hep-th/9301042].
  %%CITATION = NUPHA,B403,159;%%
}

\lref\Tong{ D.~Tong,
  ``NS5-branes, T-duality and worldsheet instantons,''
  JHEP {\bf 0207}, 013 (2002)
  [arXiv:hep-th/0204186].
  %%CITATION = HEP-TH 0204186;%%
}

\lref\GHM{R.~Gregory, J.~A.~Harvey and G.~W.~Moore,
  ``Unwinding strings and T-duality of Kaluza-Klein and H-monopoles,''
  Adv.\ Theor.\ Math.\ Phys.\  {\bf 1}, 283 (1997)
  [arXiv:hep-th/9708086].
  %%CITATION = HEP-TH 9708086;%%
}

\lref\bfss{ T.~Banks, W.~Fischler, S.~H.~Shenker and L.~Susskind,
  ``M theory as a matrix model: A conjecture,''
  Phys.\ Rev.\ D {\bf 55}, 5112 (1997)
  [arXiv:hep-th/9610043].
  %%CITATION = HEP-TH 9610043;%%
  }

\lref\milnor{J. Milnor, ``A note on curvature and fundamental group," J. Diff. Geom, 1968 p. 1-7.}

\lref\margulis{G.A. Margulis, ``Applications of Ergodic Theory to the Investigation of Manifolds of Negative
Curvature", Funct. Anal. Appl. 3 (1969), 335-336.}

\lref\Wittendual{ C.~M.~Hull and P.~K.~Townsend,
  ``Unity of superstring dualities,''
  Nucl.\ Phys.\  B {\bf 438}, 109 (1995)
  [arXiv:hep-th/9410167];
  %%CITATION = NUPHA,B438,109;%%
 E.~Witten,
  ``String theory dynamics in various dimensions,''
  Nucl.\ Phys.\ B {\bf 443}, 85 (1995)
  [arXiv:hep-th/9503124] (and generalizations).
  %%CITATION = HEP-TH 9503124;%%
}

\lref\AdSCFT{
 J.~M.~Maldacena,
  ``The large N limit of superconformal field theories and supergravity,''
  Adv.\ Theor.\ Math.\ Phys.\  {\bf 2}, 231 (1998)
  [Int.\ J.\ Theor.\ Phys.\  {\bf 38}, 1113 (1999)]
  [arXiv:hep-th/9711200];
  %%CITATION = HEP-TH 9711200;%%
  E.~Witten,
  ``Anti-de Sitter space and holography,''
  Adv.\ Theor.\ Math.\ Phys.\  {\bf 2}, 253 (1998)
  [arXiv:hep-th/9802150];
  %%CITATION = HEP-TH 9802150;%%
  S.~S.~Gubser, I.~R.~Klebanov and A.~M.~Polyakov,
  ``Gauge theory correlators from non-critical string theory,''
  Phys.\ Lett.\ B {\bf 428}, 105 (1998)
  [arXiv:hep-th/9802109].
  %%CITATION = HEP-TH 9802109;%%
  }

%\RocekPS
\lref\RV{
  M.~Rocek and E.~P.~Verlinde,
  ``Duality, quotients, and currents,''
  Nucl.\ Phys.\ B {\bf 373}, 630 (1992)
  [arXiv:hep-th/9110053].
  %%CITATION = HEP-TH 9110053;%%
}

%\BuscherSK
\lref\buscher{
  T.~H.~Buscher,
  ``A Symmetry Of The String Background Field Equations,''
  Phys.\ Lett.\ B {\bf 194}, 59 (1987);
  %%CITATION = PHLTA,B194,59;%%
T.~H.~Buscher,
  ``Path Integral Derivation Of Quantum Duality In Nonlinear Sigma Models,''
  Phys.\ Lett.\ B {\bf 201}, 466 (1988);
  %%CITATION = PHLTA,B201,466;%%
T.~H.~Buscher,
  ``Quantum Corrections And Extended Supersymmetry In New Sigma Models,''
  Phys.\ Lett.\ B {\bf 159}, 127 (1985).
  %%CITATION = PHLTA,B159,127;%%
}

\lref\Hell{
%\HellermanZM
%\lref\HellermanZM{
  S.~Hellerman,
  ``On the landscape of superstring theory in $D > 10$,''
  arXiv:hep-th/0405041;
  %%CITATION = HEP-TH/0405041;%%
%}
S.~Hellerman and I.~Swanson,
  ``Cosmological solutions of supercritical string theory,''
  arXiv:hep-th/0611317;
  %%CITATION = HEP-TH 0611317;%%
S.~Hellerman and I.~Swanson,
  ``Dimension-changing exact solutions of string theory,''
  arXiv:hep-th/0612051.
  %%CITATION = HEP-TH/0612051;%%
  }

\lref\newD{D.~Z.~Freedman, M.~Headrick and A.~Lawrence,
  ``On closed string tachyon dynamics,''
  Phys.\ Rev.\ D {\bf 73}, 066015 (2006)
  [arXiv:hep-th/0510126];
  %%CITATION = HEP-TH 0510126;%%
  T.~Suyama,
  ``Closed string tachyon condensation in supercritical strings and RG flows,''
  JHEP {\bf 0603}, 095 (2006)
  [arXiv:hep-th/0510174].
O.~Aharony and E.~Silverstein,
  ``Supercritical stability, transitions and (pseudo)tachyons,''
  Phys.\ Rev.\ D 75, 046003 (2007) [arXiv:hep-th/0612031].
  %%CITATION = HEP-TH 0612031;%%
  }

\lref\oldD{ J.~Polchinski,
  ``A two-dimensional model for quantum gravity,''
  Nucl.\ Phys.\ B {\bf 324}, 123 (1989);
  %%CITATION = NUPHA,B324,123;%%
  I.~Antoniadis, C.~Bachas, J.~R.~Ellis and D.~V.~Nanopoulos,
  ``An expanding universe in string theory,''
  Nucl.\ Phys.\ B {\bf 328}, 117 (1989);
  %%CITATION = NUPHA,B328,117;%%
A.~R.~Cooper, L.~Susskind and L.~Thorlacius,
  ``Two-dimensional quantum cosmology,''
  Nucl.\ Phys.\ B {\bf 363}, 132 (1991);
  %%CITATION = NUPHA,B363,132;%%
A.~A.~Tseytlin,
   ``Cosmological solutions with dilaton and maximally symmetric space in string
  theory,''
  Int.\ J.\ Mod.\ Phys.\ D {\bf 1}, 223 (1992)
  [arXiv:hep-th/9203033];
  %%CITATION = HEP-TH 9203033;%%
   C.~Schmidhuber and A.~A.~Tseytlin,
  ``On string cosmology and the RG flow in 2-d field theory,''
  Nucl.\ Phys.\ B {\bf 426}, 187 (1994)
  [arXiv:hep-th/9404180].
  %%CITATION = HEP-TH 9404180;%%
}

\lref\myers{
  R.~C.~Myers,
 ``New Dimensions For Old Strings,''
  Phys.\ Lett.\ B {\bf 199}, 371 (1987);
  %%CITATION = PHLTA,B199,371;%%
  %\deAlwisPR
%\lref\deAlwisPR{
  S.~P.~de Alwis, J.~Polchinski and R.~Schimmrigk,
``Heterotic Strings With Tree Level Cosmological Constant,"
  %``HETEROTIC STRINGS WITH TREE LEVEL COSMOLOGICAL CONSTANT,''
  Phys.\ Lett.\  B {\bf 218}, 449 (1989);
  %%CITATION = PHLTA,B218,449;%%
    A.~H.~Chamseddine,
  ``A Study Of Noncritical Strings In Arbitrary Dimensions,''
  Nucl.\ Phys.\  B {\bf 368}, 98 (1992).
 %%CITATION = NUPHA,B368,98;%%
%}
}

\lref\newhandle{
  A.~Saltman and E.~Silverstein,
  ``A new handle on de Sitter compactifications,''
  JHEP {\bf 0601}, 139 (2006)
  [arXiv:hep-th/0411271].
  %%CITATION = HEP-TH 0411271;%%
}

\lref\newdims{
%\McGreevyHK
%\lref\McGreevyHK{
  J.~McGreevy, E.~Silverstein and D.~Starr,
  ``New dimensions for wound strings: The modular transformation of geometry to
  topology,'', Phys.\ Rev.\  D {\bf 75}, 044025 (2007)
  [arXiv:hep-th/0612121].
  %%CITATION = HEP-TH 0612121;%%
}

\lref\mutation{
%\SilversteinQF
  E.~Silverstein,
  ``Dimensional mutation and spacelike singularities,''
  Phys.\ Rev.\ D {\bf 73}, 086004 (2006)
  [arXiv:hep-th/0510044].
  %%CITATION = HEP-TH 0510044;%%
}

%\DavidHJ
\lref\DDK{
  F.~David,
  ``Conformal Field Theories Coupled to 2D Gravity in the Conformal Gauge,''
  Mod.\ Phys.\ Lett.\  A {\bf 3}, 1651 (1988);
  %%CITATION = MPLAE,A3,1651;%%
J.~Distler and H.~Kawai,
  ``Conformal Field Theory And 2d Quantum Gravity Or Who's Afraid Of Joseph
  Liouville?,''
  Nucl.\ Phys.\  B {\bf 321}, 509 (1989).
  %%CITATION = NUPHA,B321,509;%%
}

%\ZamolodchikovWN
\lref\ZamolodchikovWN{
  A.~B.~Zamolodchikov,
  %``Infinite Additional Symmetries In Two-Dimensional Conformal Quantum Field
  %Theory,''
  Theor.\ Math.\ Phys.\  {\bf 65}, 1205 (1985)
  [Teor.\ Mat.\ Fiz.\  {\bf 65}, 347 (1985)].
  %%CITATION = TMFZA,65,347;%%
}
%\BelavinVU
\lref\BelavinVU{
  A.~A.~Belavin, A.~M.~Polyakov and A.~B.~Zamolodchikov,
  %``Infinite conformal symmetry in two-dimensional quantum field theory,''
  Nucl.\ Phys.\  B {\bf 241}, 333 (1984).
  %%CITATION = NUPHA,B241,333;%%
}

%\FreundXH
\lref\FreundXH{
  P.~G.~O.~Freund and M.~A.~Rubin,
  ``Dynamics Of Dimensional Reduction,''
  Phys.\ Lett.\  B {\bf 97}, 233 (1980).
  %%CITATION = PHLTA,B97,233;%%
}

\lref\WittenXP{
  E.~Witten and S.~T.~Yau,
  ``Connectedness of the boundary in the AdS/CFT correspondence,''
  Adv.\ Theor.\ Math.\ Phys.\  {\bf 3}, 1635 (1999)
  [arXiv:hep-th/9910245].
  %%CITATION = 00203,3,1635;%%
}

\Title{\vbox{\baselineskip12pt\hbox{NSF-KITP-07-56}
\hbox{SLAC-PUB-12439}\hbox{SU-ITP-07/05}\hbox{MIT-CTP 3829}\hbox{BRX-TH-586}\hbox{DUKE-CGTP-07-02}\hbox{UCSB Math 2007-08}}} {\vbox{ \centerline{}
%\medskip
\centerline{D{\authorfont{imensional}}-Duality } }
%The Dimensionality/Topology Correspondence
%
%
%The Incredible shrinking Riemann Surface
%
%
%
%
%
}
%\centerline{Daniel Green$^{1,2}$, Albion Lawrence$^{2,3}$, John
%McGreevy$^4$, David R. Morrison$^{2,5,6}$, and Eva Silverstein$^{1,2}$}
% alternative if we want to split the author line in half
\centerline{Daniel Green$^{1}$, Albion Lawrence$^{2}$, John
McGreevy$^3$,} \centerline{David R. Morrison$^{4,5}$, and Eva Silverstein$^{1}$}
\medskip \centerline{$^1${\it SLAC and Department of Physics,
Stanford University, Stanford, CA 94305-4060}}
%\centerline{$^2${\it
%Kavli Institute for Theoretical Physics, University of California,
%Santa Barbara, CA 93106-4030}}
\centerline{$^2${\it Brandeis Theory Group, Brandeis University, MS 057, PO Box 549110, Waltham, MA 02454}}
\centerline{$^3${\it Center for Theoretical Physics, Massachusetts Institute of Technology, Cambridge, MA
02139}} \centerline{$^4$\it{Center for Geometry and Theoretical Physics, Duke University, Durham, NC 27708}}
\centerline{$^5$\it{Departments of Mathematics and Physics, University of California, Santa Barbara, CA 93106}}
%\medskip \bigskip

\vskip.4in
%\noindent

We show that string theory on a compact negatively curved manifold,
% $\CM$,
preserving a $U(1)^{b_1}$ winding
symmetry, grows at least $b_1$ new effective dimensions as the space shrinks.  The winding currents yield a
``D-dual" description of a Riemann surface of genus $h$ in terms of its $2h$ dimensional Jacobian torus,
perturbed by a closed string tachyon arising as a potential energy term in the worldsheet sigma model. D-branes
on such negatively curved manifolds also reveal this structure, with a classical moduli space consisting of a
$b_1$-torus. In particular, we present an AdS/CFT system which offers a non-perturbative formulation of such
supercritical backgrounds. Finally, we discuss generalizations of this new string duality.

%It minimally asymptotes to a very simple background:
%a $b_1$-dimensional torus perturbed by a closed string
%tachyon, yielding a simple set of ``D-dual"
%variables appropriate for describing the small radius space.  The
%winding currents more generally translate into a description of a
%We use an analogue of Buscher's path integral derivation
%of T-duality to derive an equivalent,
%``D-dual", description of a Riemann surface
%of genus $h$ in terms of its $2h$ dimensional Jacobian torus,
%perturbed by a closed string tachyon
%arising as a potential energy term in the worldsheet sigma model.

%\draftmode

\Date{May 2007}

\newsec{Introduction}

%String and M theory dualities connect disparate backgrounds,
 %yielding enhanced control
%of physics at short distances and strong couplings.
%[We can put this back in, I just wanted to propose thinking about if we need it]
An important feature of some string dualities \Wittendual, matrix theory \bfss, and the AdS/CFT correspondence
\AdSCFT, is the emergence of new dimensions in a dual description. Previous work has focused on Ricci-flat and
positive curvature \FreundXH\ compactifications, and closely related systems. In this work, we study a new
duality with a new source of emergent dimensions, arising from compact manifolds with {\it negative}
curvature\foot{These manifolds, being much more generic than flat or positively curved spaces, could be
described as an under-represented majority of compact target spaces.}. The additional dimensions arise from the
rich topology of one-cycles in these manifolds.

In particular,  the effective central charge $c_{eff}$ is enhanced by the sum over winding sectors
\refs{\mutation,\newdims}, due to the exponential growth of the fundamental group \refs{\milnor,\margulis}.
$c_{eff}$, as a measure of the exponential growth of the density of single-string states, is a good operational
definition of dimensionality in perturbative string theory. For $c_{eff}>c_{eff}^{crit}$ the theory is
supercritical. This raises the question of whether the system admits a useful dual description in terms of a
more familiar presentation of supercritical string theory \myers.  We would like to know how large $c_{eff}$
becomes, and whether the system naturally reorganizes itself so that $c_{eff}$ arises from additional ordinary
geometrical dimensions. In this work we address this question, focusing initially on the simplest case of an
expanding Riemann surface ${\cal M}$ of genus $h$, which we find admits a natural dual description in terms of
its Jacobian torus.

In \S2\ we set up the system and indicate our regime of control.  In \S3, we introduce the symmetries of the
system and use them to derive a dual description. In \S4, we present explicit models realizing our duality,
studying them using a path integral transform along the lines of \buscher. The analysis in \S2-\S4\ pertains in
a semi-infinite range of time for which perturbative string theory applies.  In \S5\ we begin by discussing
generalizations and approaches to the initial singularity. We conclude by introducing an AdS/CFT system which
provides a candidate non-perturbative formulation of supercritical theories arising from compact hyperbolic
spaces. This makes manifest the torus corresponding to a small radius space in the infrared regime of the CFT,
and offers a concrete approach to addressing the initial singularity.
%Note it is not a Jacobian torus in this case...

\newsec{Set-up}

\lref\CHPmodels{ N.~Kaloper, J.~March-Russell, G.~D.~Starkman and M.~Trodden,
  ``Compact hyperbolic extra dimensions: Branes, Kaluza-Klein modes and
  cosmology,''
  Phys.\ Rev.\ Lett.\  {\bf 85}, 928 (2000)
  [arXiv:hep-ph/0002001];
  %%CITATION = PRLTA,85,928;%%
A.~Saltman and E.~Silverstein,
  ``A new handle on de Sitter compactifications,''
  JHEP {\bf 0601}, 139 (2006)
  [arXiv:hep-th/0411271].
  %%CITATION = JHEPA,0601,139;%%
}

We are interested in a target space containing a genus $h$ Riemann surface $\M$ as a factor.  There are many
such backgrounds, generically time-dependent, with the negative curvature driving expansion of the
space.\foot{It is also possible to consider realistic compactifications on such spaces, metastabilizing them
using extra ingredients \CHPmodels, but we will stick to simpler examples here.} It will prove particularly
simple to choose the metric at some reference time $X^0_r$ to be
\eqn\Bergmet{ds^2=\omega^a\gamma_{ab}\bar\omega^b dz d\bar z\ .}
Here $\omega^a,\bar\omega^a$ ($a=1,\dots, h$) denote the $h$ linearly independent holomorphic and
antiholomorphic 1-forms on our Riemann surface, and $\gamma_{ab}$ is a constant positive symmetric $h\times h$
matrix.  The nonlinear sigma model describing the time evolution of $\M$ is
\eqn\RSNLSM{S_{RS}= \int d^2\sigma\biggl(-(\del X^0)^2+ \omega^a(z)\partial_\nu z\gamma_{ab}\bar\omega^b(\bar
z)\partial^\nu\bar z+\delta G_{z\bar z}(X^0,z,\bar z)\del_\mu z\del^\mu\bar z+{\cal L}_\perp+h.d.\biggr)}
where ${\cal L}_\perp$ describes the dynamics of the dimensions transverse to $\M$.  Here $+h.d.$ refers to
higher derivative terms depending on the initial conditions, which are subleading at large radius but will play
a significant role at early times. The term proportional to $\delta G$ describes the time-dependent deformation
away from \Bergmet, with initial condition $\delta G(X_r^0,z,\bar x) = 0$.  Generically, the dilaton is sourced
and the worldsheet Lagrangian contains a corresponding $\int d^2\sigma \Phi(z,X^0) {\cal R}^{(2)}$ term.

In the bosonic string theory, one must tune the bare couplings in the model to cancel corrections to the
worldsheet potential (the bulk closed string tachyon).
%These corrections are small at large $R^2/\alpha'$.
We
can also study supersymmetric versions of this model, and will implement a type II GSO projection, with periodic
boundary conditions around the cycles on $\M$.  This removes bulk tachyon modes as well as tachyons from strings
winding around small homology cycles.\foot{Homologically trivial cycles still exist with antiperiodic boundary
conditions.  If these cycles become too small, they can produce transitions changing $b_0$, separating the space
into multiple components \TFA.  In this work, we will focus on the early time completion of histories where this
does not happen.}

The specific metric \Bergmet\ and the corresponding sigma model \RSNLSM\ will be convenient in formulating
explicit models and in carrying out a path integral transformation in \S4, but similar results hold more
generally.
%In the remainder of this section,
%we will discuss our framework for control of the
%time evolution and quantum effects.  In the next section, we
%will introduce the symmetries of the system, and use
%those symmetries to show that the system minimally
%asymptotes to a $T^{2h}$ at early times. These aspects
%of our analysis do not depend on the specific choice of
%metric \Bergmet.

\subsec{Time evolution and the string coupling}

One way to control the time dependence
is to study the theory in a situation where it is well-approximated by RG flow in the sigma model on $\M$ -- for
example, a supercritical theory with large spacetime dimension $D$, utilizing the timelike linear dilaton
solution with string coupling $g_s\sim g_0 e^{-\sqrt{D}X^0}$ becoming weak at late times
\refs{\oldD,\newD,\Hell,\mutation,\newdims}.\foot{Note that this large $D$ is independent of the additional
dimensions that we will find to emerge from the Riemann surface itself.} The scale dependence of operators in
the RG flow translates to the time-dependence of the associated fields, as outlined by
\refs{\DDK,\oldD,\mutation,\newD}. In particular, early times map to the UV regime of the flow, and late times
to the IR regime.   In the latter regime, RG flow is well approximated by Ricci flow.  We expect similar results
in much more general time dependent backgrounds with negatively curved spatial slices, such as the locally flat
vacuum solution considered in
\newdims, which will be useful in a candidate non-perturbative formulation presented in \S4.

%We are interested in target spacetimes with spatial slices containing a genus $h$ Riemann surface $\M$ as a
%factor. Such backgrounds are generically time-dependent, with the negative curvature driving expansion of the
%Riemann surface factor.
%Other solutions may be useful for addressing the very early time physics of the system, as we
%will discuss in \S4.  [Redundant with footnote?]
We will see that the emergence of extra dimensions in these models arises at the level of perturbative string
theory.
%and can be applied
%similarly to different weakly coupled, weakly
%curved backgrounds.  [DO WE NEED?]
We should take some care with these worldsheet arguments, since
the dilaton typically varies with time;
and for manifolds of nonconstant curvature, varies with space as well.
%the dilaton typically varies with time, as well
%as varying spatially in manifolds of nonconstant curvature.
However, in any such background the string coupling can be made arbitrarily small for any finite range of time
(over the whole compact manifold); and if the string coupling evolves monotonically, as for linear dilaton
backgrounds, the string coupling may be taken arbitrarily small over a semi-infinite range of times.\foot{In our
worldsheet path integral, we consider doing the $X^0$ integral last; a consistency check on this procedure is
the agreement with modular invariance found in \newdims.} In such backgrounds over such time intervals, our
string-tree-level discussions suffice.\foot{In the final section, we will address the very early time physics
using AdS/CFT.}

\newsec{Symmetry and duality}

The core of our duality argument is the existence of a large unbroken axial symmetry on the worldsheet, and an
associated broken vector group. String theory in $D$ dimensions compactified on $\M$ has an unbroken $U(1)^{2h}$
gauge symmetry in the $D-2$ dimensional spacetime, under which winding strings are charged.  Because the winding
number is integral, the $U(1)$s are compact.  The corresponding conserved worldsheet currents arise from the $h$
linearly independent holomorphic and antiholomorphic 1-forms $\omega^a,\bar\omega^a$ ($a=1,\dots h$) as
\eqn\axcurrent{J^a_{A\mu} \sim \epsilon_{\mu\nu}\del^\nu z\omega^a ~~~ \bar J^a_{A\mu} \sim
\epsilon_{\mu\nu}\del^\nu \bar z\bar\omega^a }
This winding symmetry, arising from $H_1\left({\cal M}\right)$, plays a key role. It is identical to the winding
symmetry of a $2h$-dimensional torus. In \S3.1\ we will show that a sigma model on $T^{2h}$ perturbed by a
relevant potential term constitutes the minimal UV completion of a Riemann surface preserving its winding
symmetries, and in \S3.2\ we will show that the winding currents translate into a description in terms of a
$T^{2h}$ much more generally. In \S4\ we will use techniques of \buscher\ to present explicit models exhibiting
the minimal completion and deformations away from it, for a Riemann surface with metric \Bergmet. In \S5\ we
will see that the same torus arises in an AdS/CFT formulation of our background.  We will call this duality --
this new set of variables appropriate to describing the small-radius surface -- ``D-duality".

\subsec{A current algebra construction of D-duality}

Nonlinear sigma models on negatively curved manifolds are infrared free and generically have Landau poles in the
UV. In our cosmological background, this corresponds to a spacelike singularity at early times, for which the
relevant degrees of freedom include a supercritical spectrum of string states \refs{\mutation,\newdims}. We
would like to understand the consistent UV completions of this model.  This is generally a difficult question.
In our case the symmetries are a powerful guide: they determine the minimal completion consistent with
maintaining them in the UV, as follows.\foot{Determining the early-time completion (related to the UV completion
of the spatial sigma model) at the level of the perturbative string theory does not in itself resolve the
spacelike singularity in situations like where the string coupling grows large at even earlier times. This
requires a non-perturbative completion, such as the AdS/CFT formulation introduced in \S4\ below.}
%
%We will see that physically, the embedding \abelmap\ of $\M$ into $\J$ provides a
%natural completion of \RSNLSM, arising in a dual formulation of the system.
%Consider a sigma model on a target space ${\cal T}=\J\times {\cal T}_\perp$ including
%a $2h$-torus $\J$, deformed by a relevant worldsheet potential term restricting the string
%to the embedded curve \abelmap\ at late times (while also lifting any transverse
%directions ${\cal T}_\perp$ in the target space). The worldsheet potential term
%corresponds in spacetime to a rolling closed string tachyon deforming the system
%away from the theory living on $\J\times {\cal T}_\perp$.
%
%\ifig\phases{Four basic classes of histories, their symmetries, and
%their effective dimensions. The flow (1) from the Jacobian to the nonlinear sigma model on ${\cal M}$ is the
%minimal completion of the latter theory which preserves the winding symmetry. Trajectories of the form (2) in
%the figure reach a greater value of the effective central charge than the minimal flow (1). In the large-radius
%regime, they differ from trajectory (1) by higher dimension operators.  Trajectories (3) involve embeddings of
%${\cal M}$ into a $c_{eff}<2h$ CFT, which do not preserve the winding symmetries. }
%{\epsfxsize2.5in\epsfbox{flow.eps}}
%
%In the next section we will discuss explicit models and show that a generalization of Buscher's path integral
%derivation of T-duality \refs{\buscher,\RV,\AAL} produces a ``D-dual" description
%with precisely this feature. [DO WE NEED?]

A {\it minimal} completion of the model will reach a fixed point in the UV (otherwise $c_{eff}$ will continue to
grow, implying arbitrarily large numbers of degrees of freedom.) We will also assume that this minimal
completion is a compact CFT (so that there are not an infinite number of winding states descending to
arbitrarily low energies), and that the target space is connected. That is, the early time background consists
of a compact CFT
%(with discrete conformal dimensions)
describing the UV fixed point of $\M$, a linear dilaton factor, and the remaining spatial directions which we
take to be flat $\IR^n$ in string frame.

If the UV completion preserves the winding symmetries, it is now straightforward to show that this minimal
completion will be string theory on ${\cal T}^{2h}$, deformed by a tachyon. The essential reason is that the
winding symmetries will be enhanced at the UV fixed point to a chiral $U(1)^{2h}\times U(1)^{2h}$ symmetry,
which can be bosonized. Consider the vertex operators for the spacetime gauge bosons coupling to the winding
charges in the timelike linear dilaton background.   The vertex operators for gauge fields polarized along
$\IR^n$ take the form $V^{\mu} = (\p Y^{\mu} \bar{J} - J \bar\p Y^{\mu}){\rm exp}(ikY)$, where $Y^{\mu}$ are free
bosons, and $J - \bar{J}$ is the current coupling to winding charge. The operators $J,\bar{J}$ must be primary
operators of dimension $(1,0)$ and $(0,1)$ respectively, for the gauge field vertex operators to be dimension
$(1,1)$. Dimension $(1,0)$ and $(0,1)$ primary operators must be chiral currents in compact CFTs\foot{This
statement has been ascribed in \refs{\ZamolodchikovWN} to \refs{\BelavinVU}, but it is not mentioned explicitly.
Here is an argument. $\bar{\p} J(z) \ket{0} = \bar L_{-1} J(z)\ket{0} \equiv \ket{\psi}$. $\brket{\psi}{\psi} =
\bra{0}J[\bar{L}_1,\bar{L}_{-1}] J \ket{0} = \bra{0}J \bar{L}_0 J\ket{0} = 0$, since $J$ has antiholomorphic
dimension 0. Therefore $\ket{\psi} = 0$ in the physical Hilbert space. Since $\ket{\psi}$ is the image under the state-operator map of
$\bar{\p} J$, we conclude that the latter must vanish.} and so generate the affine Lie algebra $U(1)^{2h}\times
U(1)^{2h}$.

These $2h$ $U(1)$ currents can be written as $\p X^k$ for $2h$ free compact bosons $X^k$. The bosons generate a
central charge $2h$, which is the order of magnitude of the contribution of the negative curvature of the
Riemann surface to the dilaton beta function, when the volume of $\M$ is of order the string scale
\refs{\mutation}. At later times\foot{Note that for simplicity, one can consider crossing over to the torus
description at a larger scale, by tuning when the tachyon turns on.  Without tuning, the tachyon condenses as
soon as it can, however.}, as the volume grows and the central charge of $\M$ flows toward $c=2$, the dual torus
must be perturbed by a relevant operator which breaks the vector subgroup of $U(1)^{2h}\times U(1)^{2h}$. This
vector subgroup is the isometry group of the torus. The most natural candidate is an $X$-dependent potential on
the worldsheet, in addition to metric perturbations which also break the isometry (and which will be generated
in any case since the tachyon appears in the metric beta functions.) We will explicitly construct examples of
these completions in the next section. In supersymmetric theories we require a description consistent with the
type II GSO projection for \RSNLSM\ -- this may add degrees of freedom, depending on the example. We will find
such completions in the next section.
%
%generate a central charge $2h$ by themselves. Since
%the total central charge of left and right movers is the same, and since our system is left-right
%symmetric on the worldsheet, this current algebra arises for both chiralities.

We conclude that a ${2h}$-dimensional torus perturbed by a tachyon is a natural worldsheet dual to a
small-radius Riemann surface which respects the winding symmetries. This torus gives the {\it minimal}
enhancement to $c_{eff}$ required to describe the UV behavior of the sigma model, given its symmetries.

Furthermore, since the winding modes in \RSNLSM\ map to the winding modes of the torus, the {\it Jacobian}
$\cal{J}$ can serve as the D-dual torus in simple examples.
%so long as $G + \delta G$ remains a complex metric on $\M$.
For any genus $h$ Riemann surface $\M$, the Jacobian is defined as a complex $h$-torus coordinatized by $X^a$,
$a=1\dots h$ with periodicity $X=X+\Omega$,
%
%\eqn\Jacdef{ds^2=d\bar X^a(Im\Omega)^{-1}_{ab}dX^b}
%
where $\Omega$ is the period matrix of $\M$. $\M$ is canonically embedded into $\cal{J}$ via the (holomorphic)
Abel--Jacobi map
\eqn\abelmap{X^a(z)=\int_{z_0}^z\omega^a\ ,}
modulo periods.  Eq.\ \abelmap\ identifies the 1-cycles of the two spaces. In the next section, we will find
simple examples in which the relevant potential -- the closed string tachyon taking the system from the minimal
$T^{2h}$ completion to the late-time Riemann surface -- realizes this canonical embedding.

This choice of torus is not mathematically arbitrary.  Any holomorphic map from ${\cal M}$ to any complex torus
${\cal T}$ must be the composition of the Abel--Jacobi \abelmap\ with an affine linear map $\J \to {\cal T}$
between the Jacobian torus $\J$ and the given torus ${\cal T}$ (see, for example, \S III.11.7 in \farkaskra).
Examples with (2,2) supersymmetry in the worldsheet matter sector will be constrained by holomorphy, in the RG
approximation to time evolution applicable at large $D$.
Such an example will be presented in \S3.2.

Note that this minimal completion includes specific massive modes at the scale of the worldsheet potential,
which translates to specific higher derivative operators in the late-time Riemann surface effective theory. More
generally, different UV completions of the sigma model Riemann surface, with different heavy modes and higher
dimension operators, may have duals with degrees of freedom in addition to the $2h$ bosons -- in other words,
the dual theory to $\M$ could have target space $\cal{J} \times \cal{T}_{\perp}$, with a potential that has $\M$
as a minimum at late times. We will find such examples in the next section; in particular the simplest way we
have found to satisfy the type II GSO projection involves a transverse sector\foot{We thank S. Hellerman for
discussions on this point.}.  In a given UV completion, the higher dimension operators may contribute small
corrections to the supercritical enhancement to the effective central charge (which was computed at leading
order in a semiclassical expansion in
\newdims).

%Achieving the minimal completion of the system consistent with the winding symmetries
%and with the %worldsheet supersymmetry applicable in the string theory of interest depends on
%a precise set of dressed higher dimension operators -- insignificant at late times -- that
%might be required on top of the na\"{\i}ve nonlinear sigma model \RSNLSM.
%In the next  section, we will derive a simple dual description of a Riemann surface model with a
%particular set of higher dimension operators.

The relation between RG flow and time evolution in our example means that the choice of a UV completion is part
of the choice of initial conditions intrinsic to all cosmological theories.  Although there may be many
consistent choices, all of them are supercritical \refs{\mutation,\newdims}, and we have learned here that those
which maintain the symmetries of the system asymptote at minimum to a $b_1$-dimensional torus at early times. In
\S4\ we will find that such a torus arises naturally in an AdS/CFT set-up, in the {\it infrared}\ regime of the
field theory side of this correspondence.
%An analog to the need to choose a good UV completion
%is what is sometimes  called the ``patch problem"
%in inflation, the requirement that the system begin in a
%smooth enough inverse-Hubble sized patch in order for inflation to set in.

\subsec{Current bosonization and duality}

The bosonization argument for D-duality, discussed in the previous subsection, generalizes to cases where we do
not reach a minimal UV completion, and to cases where RG flow and time evolution are not closely related. The
essential feature of \RSNLSM\ which leads to D-duality remains: the conserved winding currents \axcurrent\ are
{\it independent} operators in our worldsheet field theory, in the sense that there are no relations between
them in the operator algebra. This can be seen from the fact that there are $2h$ independent vertex operators
for $2h$ spacetime gauge fields coupling to the $2h$ winding modes, representing independent deformations of the
worldsheet Hamiltonian. Since the $2h$ winding currents $J^a_{A\mu}$ are conserved, as independent operators
they correspond to {\it scalar} degrees of freedom $X^a$,
\eqn\currentbos{\del_\mu J^{a\mu}_A=0 \Rightarrow J^a_{A\mu}=\epsilon_\mu^{~\nu}\del_\nu X^a}
In particular, if one wishes to work in a specific representation, their Fourier modes can be used to construct
a Fock space (which will be a subspace of the Hilbert space) equivalent to that of $2h$ scalar degrees of
freedom $X^a$, on which $J$ can be represented as in \currentbos.
%
%Furthermore, these operators $J^a_{\alpha}$ generate (via commutators or OPEs with
%other operators) $2h$ independent {\it local} $U(1)$ transformations.
%Operators creating winding string states about the $a$th cycle on $\M$
%transform nontrivially under the $a$th $U(1)$ as ${\cal O}^a_w\to e^{i\Lambda^a}{\cal O}^a_w$.
%The currents themselves transform nonlinearly under this gauged $U(1)^{2h}$. One way to see this
%directly is from the rank $2h$ Schwinger term in their OPE, which corresponds to the rank-$2h$ kinetic term for
%the corresponding gauge fields in spacetime.\foot{The Schwinger term in the commutator of two
%$U(1)$ currents $\left[J(\sigma),J(\sigma')\right] = \delta'(\sigma - \sigma')$ implies that $J$ transforms
%as $J \to J + \p\Lambda$ under local transformations generated by $J$.}
%Thus, although all of these operators are built from of a single complex field $z$, the $2h$
%$U(1)$ transformations trace out $2h$ independent local scalar functions of the worldsheet coordinates.
%
%As independent local degrees of freedom, the winding currents might be expected to play the role of
%``fundamental" fields $J^a_{A\mu}(\sigma)$ in an equivalent formulation of the system.  More precisely, since
%they are conserved, as independent fields they would correspond to scalar degrees of freedom

The fields $X^a$ typically are not free if either the vector or axial symmetries are not conserved; a generic
bosonization of $J^a_{A,\mu}$ will contain potential energy terms $T(X^a)$, corrections to the kinetic term, and
appropriate dressed higher dimension operators describing the particular history of a given model. In our models
\RSNLSM, such nontrivial interactions are required in order to reproduce the nearly critical late-time effective
central charge, and to break the symmetry under translations of $X^a$ (the dual of $J_A$.) However, none of
these corrections spoil the existence of the $2h$ independent scalar degrees of freedom.

We might also expect to construct this dual description via a generalization of Buscher's construction of
T-duality \refs{\buscher,\RV}.
%, with the winding current $J_A$ rather than the nonconserved ``momentum" currents
%$*J_A$ coupled to $2h$ auxiliary gauge fields,
%which are in turn coupled to $2h$ additional scalars.
%We will discuss such constructions in \S4.
In the next section, we consider a tractable version of this Buscher procedure, which enables us to easily
exhibit the minimal UV completion discussed in \S2, as well as more general trajectories involving the
ubiquitous $T^{2h}$. Before turning to that, we note how the symmetries affect D-brane probes of the system.

\subsec{D-branes}

%Indeed,
D-brane physics singles out the Jacobian as a possible dual of \RSNLSM.
%Recall that on a torus, the moduli space of a wrapped D-brane is the moduli space
%of WIlson lines, which coincides with the T-dual
%torus.
In the case of ordinary T-duality on a torus, the Wilson lines on wrapped D-branes trace out the T-dual torus.
On a compact negatively curved  space ${\cal M}_n$ of any dimension $n$ and first homology $b_1$, a wrapped
D-brane has as its classical moduli space of Wilson lines a torus $T^{b_1}$ (up to curvature couplings that drop
out in the low-energy Yang-Mills limit, and in the case of a flat spacetime solution as considered in \newdims).
Note that this torus will be the T-dual of the Jacobian -- the Abel-Jacobi map sends $H_1({\cal M}_n) \to
H_1(\cal{J})$.

In \S5\ we will present an AdS/CFT system comprised of many D-branes which manifests a $T^{b_1}$ in the regime
corresponding to the {\it IR}\ physics of the field theory. This last construction may address the very early
time physics in the solutions to which it applies.

\newsec{Explicit construction of dual pairs}

\lref\MPtowards{
  D.~R.~Morrison and M.~R.~Plesser,
  ``Towards mirror symmetry as duality for two dimensional abelian gauge
  theories,''
  Nucl.\ Phys.\ Proc.\ Suppl.\  {\bf 46}, 177 (1996)
  [arXiv:hep-th/9508107].
  %%CITATION = NUPHZ,46,177;%%
}

\subsec{Generalizations of T-duality}

In constructing explicit D-duals, we follow a variant of the basic strategy used by Buscher \refs{\buscher,\RV}\
to derive T-duality via the path integral: that is, couple each {\it winding} current\foot{In essence Buscher's
construction is a gauging of the vector symmetry whose conserved charge is target space momentum.} to a gauge
field,  and add Lagrange multiplers $X^a$ (which will become the same periodic scalars $X^a$ appearing in the
bosonization discussions in \S3) that keep the gauge field trivial.  Beginning with this theory, gauge fixing
and integrating out trivial degrees of freedom in different orders leads to different presentations of identical
quantum theories.
%
%A generalization of Buscher's path integral derivation of T-duality \refs{\buscher,\RV}\ suggests the same
%result.  One can gauge the winding symmetry on the worldsheet, and introduce $2h$ Lagrange multipliers $X^a,\bar
%X^a$ to force the gauge field to be trivial, ensuring that the model is equivalent to the original. Integrating
%out the gauge field first instead yields a description in terms of $X^a$.
As in generalizations of T-duality and mirror symmetry found in \eg\ \refs{\MPtowards,\OV,\GHM,\Tong,\HV}, one
finds that a potential $T(X^a)$ and other corrections are allowed in the dual.

In the subsections below, we will consider a tractable version of this Buscher procedure (for theories with
zero, two, and four supercharges) which enables us to easily exhibit the minimal completion discussed in \S3, as
well as more general trajectories involving a $T^{2h}$.  These involve massive degrees of freedom that are part
of the UV completion of the theory, which reduce to specific higher derivative terms in \RSNLSM\ at late times.
%There may be constructions closer to the original
%version of \refs{\Buscher,\RV} which have fewer degrees of
%freedom in the UV.  We leave a study of these for future work.
%
%In formulating this, we encountered difficulties with Wick rotation which
%prevent us from carrying through the
%transformation explicitly.  In any case, its essence appears to be the
%simple relation \currentbos.
%
%In the next subsection, we consider a tractable variant of this Buscher procedure, which enables us to easily
%exhibit the minimal UV completion discussed in \S2, as well as more general trajectories involving the
%ubiquitous $T^{2h}$.

\subsec{Bosonic dual pair}

We will begin with a purely bosonic worldsheet theory -- this will capture the qualitative features of the
duality. In \S4.3\ below we will add ${\CN}=(1,1)$ worldsheet supersymmetry in order to rid ourselves of
additional tachyons that would destabilize the late-time behavior of the model.  In \S4.4\ we will consider an
$\CN=(2,2)$ version.

%In terms of \RSNLSM\ we will be studying the
%duality at fixed reference time $X^0_i$. In cases where the time
%dependence is slow, such as when we
%consider the reduction of a highly supercritical string theory on $\M_h$, we
%can consider this duality fiberwise.  In principle we will need to take care with the cutoff of the theory,
%since it is then only conformal upon dressing it with $X^0$.
%[We already said these things in the setup section -- ES]

We consider the bosonic theory with action (at fixed reference time $X_r^0$)
\eqn\Sbos{\eqalign{{\cal S}=\int d^2\sigma\gamma_{ab} \biggl\{& (\del Y^a-A^a)(\del\bar Y^b-\bar A^b)^2-i(X^a
\bar F^b+F^a\bar X^b)\cr & +i\left[\tilde F^a(\bar X^b-\int^{\bar
z}\bar\omega^b)+(X^a-\int^z\omega^a)\bar{\tilde F}^b\right]+{1\over e^2}\tilde F^a\bar{\tilde F^b}\biggr\}}}
where $\tilde F$ is the field strength for a complex gauge field $\tilde A$, and similarly $F=\del_{\mu}A_{\nu}
\epsilon^{\mu\nu}$. As stated above, we can we can consider different combinations of gauge fixing and
integrating out trivial degrees of freedom; all theories which are derived in this way from \Sbos\ should be
equivalent as quantum theories. We will look at two particular such sequences.  The first sequence leads to a
sigma model with variable $z$ whose target space is the Riemann surface $\M_h$ (plus some higher-derivative
terms corresponding to massive degrees of freedom); the other of which gives us a sigma model on a
$2h$-dimensional torus, together with a nontrivial potential. Note that in \Sbos, the gauge field $\tilde A$
couples to the axial (winding) currents of {\it both} $z$ and $X$. In this sense \Sbos\ is a variant of the
usual Buscher action: in \refs{\buscher,\RV}, one couples the auxiliary gauge field to the target space momentum
current $\del z$ of one variable, and the target space winding current $*\del X$ of the dual variable.

 %Note also the gauge kinetic term in \Sbos\  for $\tilde F$.
 %The coupling $e^2$ introduces a new scale which,
 %once dressed with appropriate time dependence, determines
 %when the system crosses over from a simpler UV
 %description to the Riemann surface nonlinear sigma model.

 %is in principle independent of the UV cutoff
 %needed to define \Sbos.

Consider first integrating out $X^a$, which appears as a Lagrange multiplier. This forces $F=\tilde F$.  If we
then fix gauge by setting $Y=0=A-\tilde A$ we find a theory of massive vectors (of mass $e$) coupled to $z$:
\eqn\Azac{\int d^2\sigma\gamma_{ab}\biggl\{A^a\bar A^b+i\left[F^a\int^{\bar z}\bar\omega^b+\int^z\omega^a\bar
F^b\right]+{1\over e^2}F^a\bar{F^b} \biggr\} }
At energies below the scale $e$ we integrate out $A$ to find:
\eqn\SbosRS{{\cal S}\sim\int d^2\sigma \biggl\{\omega^a\del z\gamma_{ab}\bar\omega^b\del\bar z+h.d.\biggr\}}
where ``$h.d.$'' stands for higher derivative terms suppressed by powers of $\del^2/e^2$.  Our model thus
produces a particular UV completion of \RSNLSM.  At low energies it is the two-derivative nonlinear sigma model
on our Riemann surface, together with a specific set of higher dimension operators which are irrelevant at large
radius. At the scale $e$ (which can be taken to be lower than the strong coupling scale of the Riemann surface
nonlinear sigma model) we find additional degrees of freedom in the form of massive vector fields. The finite
mass of the vector fields follows from the gauge kinetic term for $\tilde F$ in \Sbos.\foot{If we gave the gauge
field in
 \refs{\buscher,\RV}\ a kinetic term we would find that this term is a redundant operator.
 In the present case it is not -- it is part of the definition of the UV completion of the theory.}

Let us now integrate in a different order to obtain a second description which makes manifest the D-dual
variables appropriate to early times (equivalently, scales above $e$). Gauge fixing $Y = 0$ and integrating over
$A$ leads to the theory
\eqn\Sbosdual{\tilde{\cal S}=\int d^2\sigma\gamma_{ab}\biggl\{ \del X^a\del\bar X^b+i\left[\tilde F^a(\bar
X^b-\int^{\bar z}\bar\omega^b)+(X^a-\int^z\omega^a)\bar{\tilde F}^b\right]  +{1\over e^2}\tilde F^a\bar{\tilde
F^b} \biggr\}\ .}
The gauge fields $\tilde A$ couple to winding currents and do not propagate.  $X - \int^z \omega$ couples as an
axion to $\tilde F$. As explained in \refs{\Coleman,\Edphases}, this theory has an effective axion potential
coming from the electric field, given at one loop, for $|X^a-\int^z\omega^a|<\pi$, by
\eqn\CW{{\cal U}=e^2(X^a-\int^z\omega^a)\gamma_{ab}(\bar X^b-\int^{\bar z}\bar\omega^b)}
Note that we expect the full axion potential to be periodic since \Sbosdual\ has this periodicity (due to flux
quantization). This should arise from the pair production of winding states, as explained in \Coleman.

In \CW, $z$ is non-propagating and we can integrate it out.  This will lead to an effective potential $V(X)$ for
$X$. If we integrate out $z$ classically, then $V(X)$ will vanish whenever $X$ takes a value for which ${\cal
U}$ vanishes; this occurs when $X$ is at an image point of the Abel-Jacobi map. Therefore, Eq.\ \CW\ provides
precisely the expected potential restricting the system to the Riemann surface. Note that under this map, $*\del
X$ and $*\omega\del z$ are interchanged; the duality maps winding states in the Riemann surface into those of
the Jacobian.

Upon fibering \Sbos\ over the time direction, $V(X)$ behaves as a spacetime tachyon and should grow at late
times.  At these late times, \SbosRS\ will be a better description, while at early times the description in
terms of the D-dual variables $X$ becomes appropriate since $V(X)$ becomes smaller and smaller. That is, we have
found a model realizing the minimal early time completion determined on general grounds in \S3.2. The path
integral transform itself is fairly trivial; note that in the absence of the $z$ fields, it simply relates
massive vectors $A^a$ on the first side to the equivalent theory of massive scalars $X^a$ on the second side,
and both sides manifestly reduce to a Riemann surface at low energies.

To describe trajectories different from the minimal completion of \S2.3, we may formally add or subtract higher
dimension operators from both sides.  For example, if we subtract the $+h.d.$ terms from \SbosRS, we obtain, at
a given timeslice, the basic nonlinear sigma model on a Riemann surface.  By carrying through the duality
transformation to the $X$ variables, one subtracts the corresponding sum of $h.d.$ terms, now expressed in terms
of $X$ (using the fact that these higher dimension operators are all functionals of the winding current and its
derivatives, which translates into $*\del X$ and its derivatives). We do not know a priori if this deformed
theory in itself is UV complete, but if so we see from this transformation that it also has a dual description
in terms of coordinates $X$ on its Jacobian torus. Similar remarks apply to other trajectories obtained by
deforming \SbosRS.  This reflects the genericity of the bosonization \currentbos.

\subsec{An $\CN=(1,1)$ supersymmetric extension}

We now wish to construct a $\CN=(1,1)$ supersymmetric version of \Sbos, in which there is no spacetime tachyon
destabilizing \RSNLSM\ at late times.

We follow the conventions of \susycon.  Consider the type II version on \RSNLSM. The Riemann surface directions
(at a fixed reference time $X^0_r$) can be described by a (1,1) supersymmetric nonlinear sigma model involving
scalar superfields $Z,\bar Z$:
\eqn\IIRS{{\cal S}_{II}=\int d^2\sigma d^2\theta \gamma_{ab}\biggl\{\omega^a D_\alpha Z\bar\omega^b D^\alpha\bar
Z+h.d.\biggl\}}
where again ``$h.d.$" stands for higher derivative operators, subleading at large radius.  This theory respects
a $\IZ_2\times\IZ_2$ R-symmetry by which we can orbifold to implement the type II GSO projection, giving a
tachyon-free spectrum.\foot{The time evolution produces harmless {\it pseudotachyonic} modes of the massless
fields, as discussed recently in \refs{\newdims, \Hell,\newD}.}  In perturbative string theory, we can choose
any spin structure on $\M$. The fully periodic spin structure avoids the decays of $b_1$ due to winding tachyons
that can occur when $\M$ is small \TFA.

The generalization of \Sbos\ to (1,1) supersymmetry is %%
%\eqn\Ssusy{\eqalign{S= \int d^2\sigma d^2\theta & \gamma_{ab}\biggl\{ \left(D_\alpha
%Y^a-\Gamma^a_\alpha\right)\left(D^\alpha\bar Y^b-\bar\Gamma^{b\alpha}\right) -i[X^a
%D^\alpha(\sigma^3\bar\Gamma)^b_\alpha+\bar X^b D^\alpha(\sigma^3\Gamma)^a_\alpha]\cr & + i[(X^a-\int^Z\omega^a)
%D^\alpha(\sigma^3\bar{\tilde\Gamma})^b_\alpha+(\bar X^b-\int^{\bar Z}\bar\omega^b)
%D^\alpha(\sigma^3\tilde\Gamma)^a_\alpha]\cr &+{1\over {4e^2}}(D^\beta D_\alpha\tilde\Gamma^a_\beta)(D^\delta
%D_\alpha\bar{\tilde\Gamma}^b_\delta) \biggr\} \cr}}
%%
%
%\eqn\Ssusy{\eqalign{ S= \int d^2\sigma d^2\theta & \gamma_{ab}\left\{ \left(D_\alpha
%Y^a-\Gamma^a_\alpha\right)\left(D^\alpha\bar Y^b-\bar\Gamma^{b\alpha}\right) -i[X^a \bar f^b +
%%D^\alpha(\sigma^3\bar\Gamma)^b_\alpha+
%\bar X^b f^a]
%%D^\alpha(\sigma^3\Gamma)^a_\alpha]\cr &
%\right. \cr & \left. + i \left[ \left(X^a-\int^Z\omega^a \right) \bar{\tilde f^b}
%%D^\alpha(\sigma^3\bar{\tilde\Gamma})^b_\alpha+
%+\left(\bar X^b-\int^{\bar Z}\bar\omega^b \right) \tilde f^a \right]
%%D^\alpha(\sigma^3\tilde\Gamma)^a_\alpha]\cr &
%+{1\over {4e^2}} \tilde f^a \bar{\tilde f^b}
%%(D^\beta D_\alpha\tilde\Gamma^a_\beta)(D^\delta
%%D_\alpha\bar{\tilde\Gamma}^b_\delta) \biggr\} \cr}}
%\right\} }}
%
% v2:
\eqn\Ssusy{\eqalign{ S= &\int d^2\sigma d^2\theta  \gamma_{ab}\left\{ \left(D_\alpha
Y^a-\Gamma^a_\alpha\right)\left(D^\alpha\bar Y^b-\bar\Gamma^{b\alpha}\right) -i[X^a \bar f^b +
%D^\alpha(\sigma^3\bar\Gamma)^b_\alpha+
\bar X^b f^a]
%D^\alpha(\sigma^3\Gamma)^a_\alpha]\cr &
\right. \cr & \left. + i \left[ \left(X^a-\int^Z\omega^a \right) \bar{\tilde f^b}
%D^\alpha(\sigma^3\bar{\tilde\Gamma})^b_\alpha+
+\left(\bar X^b-\int^{\bar Z}\bar\omega^b \right) \tilde f^a \right]
%D^\alpha(\sigma^3\tilde\Gamma)^a_\alpha]\cr &
+{1\over {2e^2}}  % \tilde f^a \bar{\tilde f^b}
(D^\beta D_\alpha\tilde\Gamma^a_\beta)(D^\delta D^\alpha\bar{\tilde\Gamma}^b_\delta)
%\biggr\} \cr}}
\right\} }}
Here $D_\alpha$ is the superspace derivative, and $Y^a, X^a $ and $Z^a$ are complex scalar multiplets.
$\Gamma^a,\tilde\Gamma^a$ are gauge multiplets, whose scalar field strengths are $f^a \equiv
%The matrix $\sigma^3$ appearing in the axion terms gives
D^\alpha(\sigma^3\Gamma^a)_\alpha = D_+\Gamma^a_- + D_-\Gamma^a_+$; here $\pm$ indicates the helicity in
multiples of 1/2 as described in \susycon. $\Gamma$ and $\tilde\Gamma$ flip sign under each of the chiral
$\IZ_2$ R-symmetries, while the scalar multiplets are neutral.

To find a presentation of the quantum theory in which $\M$ is manifest, we first integrate out $X$, yielding
$\Gamma=\tilde\Gamma+D\Lambda$.
Fixing the gauge symmetries associated with $\Gamma$ and $\tilde\Gamma$ can be done most simply by setting
$Y=0=\Lambda$.  Next, integrating out $\Gamma$ gives
\eqn\firstreln{\Gamma-DY=D\int^Z\omega+{\cal O}(\del^2/e^2)}
which yields the Riemann surface sigma model \IIRS, with the higher derivative terms arranged in a specific
power series in $\del^2/e^2$.

In order to find a presentation of the quantum theory in which $T^{2h}$ is manifest, we begin by integrating out
$\Gamma$, which we can do by solving $\Gamma-DY+DX=0$ for $\Gamma$.
%
%One component of the superspace expansion of \gammaeom\ is the equation $A-*\del
%Y+*\del X=0$, seen in the bosonic case.
%\foot{Note that combining the superspace relation
%with \firstreln\ and the gauge-fixing setting $Y=\Lambda=0$ leads to an
%identification of winding modes on the two sides of the duality.}
Plugging this into Eq. \Ssusy, we obtain %%
%\eqn\Sdual{\eqalign{\tilde S=\int d^2\sigma d^2\theta\gamma_{ab}\biggl\{ & D_\alpha X^a D^\alpha \bar X^b +
%i[(X^a-\int^Z\omega^a) D^\alpha(\sigma^3\bar{\tilde\Gamma})^b_\alpha+(\bar X^b-\int^{\bar Z}\bar\omega^b)
%D^\alpha(\sigma^3\tilde\Gamma)^a_\alpha]\cr & +{1\over {4e^2}}(D^\beta D_\alpha\tilde\Gamma^a_\beta)(D^\delta
%D_\alpha\bar{\tilde\Gamma}^b_\delta) \biggr\}}}
%
\eqn\Ssusy{\eqalign{ S= \int d^2\sigma d^2\theta & \gamma_{ab}\left\{ D_\alpha X^a D^\alpha \bar X^b
%\left(D_\alpha
%Y^a-\Gamma^a_\alpha\right)\left(D^\alpha\bar Y^b-\bar\Gamma^{b\alpha}\right)
%-i[X^a \bar f^b +
%%D^\alpha(\sigma^3\bar\Gamma)^b_\alpha+
%\bar X^b f^a]
%%D^\alpha(\sigma^3\Gamma)^a_\alpha]\cr &
+ i \left[ \left(X^a-\int^Z\omega^a \right) \bar{\tilde f^b}
%D^\alpha(\sigma^3\bar{\tilde\Gamma})^b_\alpha+
+ h.c. \right]
%
%+\left(\bar X^b-\int^{\bar Z}\bar\omega^b \right) \tilde f^a \right]
%
%D^\alpha(\sigma^3\tilde\Gamma)^a_\alpha]\cr &
%v2:
+{1\over {2e^2}}
%\tilde f^a \bar{\tilde f^b}
(D^\beta D_\alpha\tilde\Gamma^a_\beta)(D^\delta D^\alpha\bar{\tilde\Gamma}^b_\delta)
%\biggr\} \cr}}
\right\} }}

As in the bosonic case discussed above, the electric field energy translates \'a la \Coleman\ into a potential
term \CW\ for the scalar field components. There are additional terms in the supersymmetric action, involving
the canonically normalized scalar $\tilde a_3$ in the $\tilde\Gamma$ gauge multiplet:
\eqn\Ususy{{\cal U}=e^2(X^a-\int^z\omega^a)\gamma_{ab}(\bar X^b-\int^{\bar z}\bar\omega^b)+e^2\gamma_{ab}\tilde
a_3^a\bar{\tilde a_3}^b + e\gamma_{ab}\left(\bar F_Z\bar\omega^a\tilde a_3^b + {\bar{\tilde
a}_3}^aF_Z\omega^b\right) , }
where $F_Z$ is the auxiliary field in the $Z$ multiplet. Note that the scalar $a_3$ flips sign under the chiral
GSO projection -- this fact allows for a scalar potential consistent with a GSO projection, giving a stable
late-time solution. This is reminiscent of a mechanism in \Hell\ for obtaining dimension-changing transitions
yielding the critical type II theory on an orbifold fixed locus.\foot{Indeed, the existence of an example with
some of these features was anticipated by S. Hellerman.}
% v2:
The integral over $F_Z$ constrains a linear combination of $\tilde a_3$s
to vanish identically.
In this theory $Z$ is heavier than $X$ and the remaining $\tilde
a_3$s, and can be integrated out.  Although it is difficult to carry out this path integral in practice, the form
of the potential shows that at late times (when $e^2$ is large), it yields a theory with a tachyon which
restricts the $X$ and $\tilde a_3$ directions to the embedded Riemann surface.  This matches the low energy
physics of the Riemann surface model \IIRS.

The GSO projection also protects the model against arbitrary tachyon deformations at early times.  Since $\tilde
f^a$ transforms under the GSO and $X^a$ and $Z$ do not, the only potential terms allowed for $X$ and $Z$ are
those coming from the electric field energy arising from the axion couplings of the form $i(\bar{\tilde
f}^a\gamma_{ab}\theta^b(X,Z)+h.c.)$. In general, this yields only $h$ complex constraints on the $h+1$ complex
variables $X^a$ and $Z$, so the low energy theory is a Riemann surface generically (without fine tuning).

More specifically, we are considering a Riemann surface which maintains the winding symmetries of the torus.
This is at least a self-consistent symmetry principle in the RG evolution of the system applicable at large $D$.
It is worth noting also that a gas of winding strings would energetically favor the subset of tachyons
preserving the winding symmetries of the torus, since tachyons which remove the corresponding cycles would
render winding strings heavy.

\subsec{An $\CN=(2,2)$ supersymmetric extension}

At fixed reference time $X^0_r$, the (1,1)-supersymmetric model just discussed has a simple extension with (2,2)
supersymmetry. Since $\CN=(2,2)$ SUSY implies a complex target space, the time dependence will generally break
$N=(2,2)$ supersymmetry.  On the other hand, if we consider the expanding Riemann surface as a factor in a
highly supercritical theory in $D+10$ dimensions, the time dependence will be slow, and we can think of the
cosmology as a trajectory for an $\CN=(2,2)$ sigma model, with supersymmetry broken to $(1,1)$ by corrections of
order $1/\sqrt{D}$. There are two advantages to this presentation. First, the complex target space means that
the D-duality will involve a holomorphic map of $\M$ into $T^{2h}$, for which there is a natural set of
mathematical objects discussed in \S3.1. Secondly, gauging a symmetry in $(2,2)$ language removes two scalar
degrees of freedom -- one from the gauge fixing at low energies, and one from the D-term constraints. Thus, we
need half of the gauge fields of the $\CN=(1,1)$ and $\CN=(0,0)$ cases, to realize the duality.

The field content of the $\CN=(2,2)$ model is: $h$ twisted chiral multiplets $X^a$, $h$ chiral multiplets $Y^a$,
one twisted chiral multiplet $Z$ which lives on $\M$, and two sets of $h$ vector superfields $V,\tilde{V}$ with
twisted chiral field strengths $\Sigma^a = D_+ \bar{D}_- V$, $\tilde{\Sigma}^a = D_+ \bar{D}_- \tilde{V}$. The
$\CN=(2,2)$ Lagrangian (corresponding to the $(2,2)$ completion of \RSNLSM\ at fixed reference time $X^0_i$) is
%, suppressing the $b_1$ indices
%(which are all contracted with $\gamma_{ab}$ as usual), is
%$$ \int d^2\tilde \theta
%\left[ \tilde \Sigma ( X - \int^Z \omega) - \Sigma X \right] + {\rm h.c.}
%+ \int d^4 \theta\left[ ( Y + \bar Y + V)^2 +
%\tilde \Sigma^\dagger \tilde \Sigma/e^2 \right] . $$
%
\eqn\twotwo{ \eqalign{ L = & \int d^2\tilde \theta \half \gamma_{ab}  \left[ \tilde{\bar{\Sigma}}^a \left( X^b -
\int^Z \omega^b \right) - {\bar{\Sigma}}^a X^b \right] + {\rm h.c.} \cr & + \int d^4 \theta \gamma_{ab} \left[
\half( Y^a + \bar Y^a + V^a)( Y^b + \bar Y^b + V^b) - {1\over 2e^2}\tilde \Sigma^{a~\dagger} \tilde \Sigma^b
\right]\ . }}
Here $\int d^2 \tilde\theta \equiv \bar D_+ D_-$.
%Note that the (2,2) model is in a sense more economical in
%that there only $2h$ gauge multiplets, as opposed to $4h$ in the (1,1) case.  The reason for this is that the
%potential \CW\ will arise in (2,2) supersymmetry from a combination of the electric field and the $D$ term
%energy.

To get the (2,2) Riemann surface model, we integrate out $X$ first, so that $ \Sigma = \tilde \Sigma $ up to a
(2,2) supergauge transformation which we gauge fix to zero.  We also choose the gauge $Y=0$. The remaining
lagrangian, suppressing the $b_1$ indices which are always contracted with $\gamma_{ab}$, is
\eqn\backtoRS{L(V,Z)= \int d^4 \theta ~\left\{\half V^2 + \CO\left(e^{-2} \right) \right\} - \half \int
d^2\tilde\theta ~\Sigma \int^Z \omega + {\rm h.c.}. } Consider the twisted superpotential term in \backtoRS.
Since $\int^Z \omega$ is twisted chiral, this term is a total superspace derivative: $ \half \int
d^2\tilde\theta ~\Sigma \int^Z \omega = \int d^4 \theta ~V \int^Z \omega . $ The resulting action is:
\eqn\VZlag{ L(V,Z)=\int d^4 \theta \left\{ \half V^2 - V \left(\int ^Z \omega + {\rm h.c.} \right)+ \CO(e^{-2})
\right\}. } Finally, integrating out $V$ produces the lagrangian (up to higher derivative terms at scale $e$,
and a k\"ahler transformation) \eqn\kahlerRS{ L_{\rm RS}(Z)=- \int d^4 \theta \left|\int^Z \omega \right|^2 }
which describes the K\"ahler sigma model on the Riemann surface\foot{Note that this sign gives the correct
kinetic terms for {\it twisted} chiral fields.}.

Performing the integrations in a different order produces a $(2,2)$ version of the D-dual, with the manifest
$T^{2h}$. Let us begin by rewriting \twotwo\ as: \eqn\fullsuperspaceform{ L = \int d^4 \theta \left[ \half(V + Y
+ \bar Y)^2 - V (X + \bar X) + \tilde V \left( X - \int^Z \omega + {\rm h.c.} \right) - {1\over 2e^2}\tilde
\Sigma^\dagger \tilde \Sigma \right] } Integrating out gauge multiplet $V$ relates the $X$ current to the $Y$
axial current
%and $\omega \del z$ currents
as usual in (2,2) duality \refs{\eg\ \HV}: \eqn\Veom{  (V + Y + \bar Y) - (X + \bar X) = 0 .} (From this
relation, we see that $Y$ is the coordinate on the {\it mirror} of the Jacobian, the torus related to the
Jacobian by T-duality on half the circles.)

The remaining Lagrangian is \eqn\XZtildeVlag{ L| =\int d^4 \theta \left[ - {1\over 2} (X + \bar X)^2 - \tilde V
(t +\bar t) -{1\over 2e^2}\tilde \Sigma^\dagger \tilde \Sigma \right] } where $ t \equiv X - \int^Z \omega $ The
dynamics of the tilded gauge multiplet generates a potential for its theta angle $t$, minimized at zero
\Edphases.  To see this explicitly, rewrite the
%$\tilde V$-dependent terms in the
lagrangian as \eqn\sigmaXzlag{ L|=\int d^4 \theta ~ \left(- {1\over 2 e^2} \tilde \Sigma^\dagger \tilde \Sigma -
X^\dagger X \right) + \half \int d^2 \tilde \theta \tilde \Sigma t + {\rm h.c.} } which is effectively a
Wess-Zumino model for the twisted chiral fields $\tilde \Sigma, X, Z$, with twisted chiral superpotential $
\tilde W = \tilde \Sigma t $ (the $\tilde V$ gauge symmetry doesn't act on anything and is trivially confined).
Integrating out the auxiliary fields in the twisted chiral multiplets $X, \tilde \Sigma$, and rescaling
$\tilde\sigma$ to have canonical kinetic term, the bosonic potential
%is the usual
%\eqn\usualVbos{ g^{ab} {\del \tilde W \over \del \tilde \Sigma^a } \overline{ {\del\tilde W \over \del \tilde \Sigma^b } } }
%which
is \eqn\CWpotential{ V(X, z, F_z) = e^2 \left|  X - \int^z \omega \right|^2 + e^2\left| \tilde \sigma \right|^2
+ \left( e\tilde \sigma \cdot \omega F_z + {\rm h.c.} \right) . }
%e^2 \gamma_{ab} \left( X^a - \int^z \omega^a \right)
%\overline{ \left( X^b - \int^z \omega^b \right) } + \gamma_{ab} \tilde \sigma^a
%\bar{\tilde \sigma}^b.}
This is the potential \CW, plus a mass for the scalars $\tilde \sigma$. The integral over the auxiliary field
$F_z$ imposes a constraint $ \sum_a \omega^a(z) \tilde \sigma^a = 0 $.
%This condition wants to imply that there are $2h-2$ other vacua of the sigmas
%which makes the witten index match between IR and UV.

As in the bosonic and (1,1) models, integrating out $z$ now generates a potential for $X$ which vanishes when
$X$ lies in
%This potential restricts the field configurations to
the image of the
%Riemann surface by the
Abel-Jacobi map.

\newsec{Discussion}

%\ArkaniHamedMZ
\lref\axionpot{
%\HosotaniXW
  Y.~Hosotani,
  ``Dynamical Mass Generation By Compact Extra Dimensions,''
  Phys.\ Lett.\  B {\bf 126}, 309 (1983);
  %%CITATION = PHLTA,B126,309;%%
%\HatanakaYP
  H.~Hatanaka, T.~Inami and C.~S.~Lim,
  ``The gauge hierarchy problem and higher dimensional gauge theories,''
  Mod.\ Phys.\ Lett.\  A {\bf 13}, 2601 (1998)
  [arXiv:hep-th/9805067];
  %%CITATION = MPLAE,A13,2601;%%
%\AntoniadisCV
  I.~Antoniadis, K.~Benakli and M.~Quiros,
  ``Finite Higgs mass without supersymmetry,''
  New J.\ Phys.\  {\bf 3}, 20 (2001)
  [arXiv:hep-th/0108005].
  %%CITATION = NJOPF,3,20;%%
%\vonGersdorffAS
  G.~von Gersdorff, N.~Irges and M.~Quiros,
  %``Bulk and brane radiative effects in gauge theories on orbifolds,''
  Nucl.\ Phys.\  B {\bf 635}, 127 (2002)
  [arXiv:hep-th/0204223];
  %%CITATION = NUPHA,B635,127;%%
%\ChengIZ
  H.~C.~Cheng, K.~T.~Matchev and M.~Schmaltz,
  ``Radiative corrections to Kaluza-Klein masses,''
  Phys.\ Rev.\  D {\bf 66}, 036005 (2002)
  [arXiv:hep-ph/0204342].
  %%CITATION = PHRVA,D66,036005;%%
  N.~Arkani-Hamed, H.~C.~Cheng, P.~Creminelli and L.~Randall,
  ``Pseudonatural inflation,''
  JCAP {\bf 0307}, 003 (2003)
  [arXiv:hep-th/0302034].
  %%CITATION = JCAPA,0307,003;%%
}

In this final section we briefly discuss perturbative
generalizations of D-duality, and then
introduce a possible non-perturbative completion using AdS/CFT.

\subsec{Generalizations and transitions}

Our construction of D-duality should work equally well for higher-dimensional target spaces $\M$ with negative
curvature. For negatively curved manifolds with $b_1 > 0$, the current algebra arguments apply at early times to
show that the minimal completion is a $T^{b_1}$. Mathematically, there are natural candidate D-duals. For
higher-dimensional K\"ahler manifolds $\CM$, the analogue of the Jacobian torus $\cal{J}$ of $\CM$ is known as
the ``Albanese variety'', and the analogue of \abelmap\ is known as the ``Albanese map". In such cases,
holomorphic maps from $\M$ to a complex torus $\cal{T}$ should be the composition of the Albanese map and map
from the Albanese variety to $\cal{T}$. More generally, when the target manifold is not complex, we might expect
a similar structure to hold for harmonic maps, relating winding modes on $\M$ to those on the associated
$T^{b_1}$.

Nontrivial $c_{eff} - c_{crit}$ arises essentially from the exponential growth of the first homotopy group
$\pi_1(\M)$, following from the compactness and negative curvature of our target space
\refs{\mutation,\newdims}. In the present work, the first homology $H_1(\M)$ plays a key role. However, there
exist negatively curved compact manifolds which are supercritical by virtue of the exponential growth of
$\pi_1$, but which have $b_1=0$.  This raises the question of what if any D-dual set of variables describes
their small volume limit.
%Therefore, we may also be able to generalize this
%duality to negatively curved manifolds with $b_1 = 0$. For
%example,
Recall that in T-duality, a similar question arises for orbifolds which preserve neither winding nor momentum
symmetry.  The duality does extend to orbifolds, which inherit T-duality properties of the parent theory defined
on the covering space. For similar reasons, we expect that our derivation will generalize to yield a dual
description of any space ${\cal M}_n$ in terms of a $\tilde b_1$-dimensional torus, where $\tilde b_1$ is the
first Betti number of any finite cover of ${\cal M}_n$ (known as a {\it virtual} first betti number of $\M_n$).
For example, there exist many hyperbolic $3$-manifolds which are {\it homology spheres}, i.e., $H_1=\{0\}$ so
that $b_1=0$ \kawauchi. Mathematicians have conjectured that these always have finite covers with nonzero
$\tilde b_1$ (see, for example, \masters); this would provide a way to extend D-duality to general 3-manifolds
with $c_{eff}
> c_{crit}$. Conversely, it would be interesting to see if our physical methods can provide insight into the
validity of the virtual first betti number conjecture.

String theory contains dimension-changing transitions (see \refs{\newD,\Hell}\ for
recent well-controlled examples).  Having learned here and in \refs{\mutation,\newdims}\ that
string theory on negatively curved target spaces is supercritical, it is
interesting to consider transitions to lower dimensions (equivalently,
lower $c_{eff}$) in this context.  The expansion of the space itself
yields such a transition, since $c_{eff}$ decreases in
time. The results in \TFA\ yield another decay mode in some cases (with spin
structure chosen to be antiperiodic for spacetime fermions about some handle(s)),
in which the reduction in genus reduces $c_{eff} \propto \frac{2h-2}{V_{\Sigma}}$.

\subsec{Early times}

The duality obtained above is essentially perturbative, and can be studied during a semi-infinite
epoch of arbitrarily weak string coupling.  In the timelike linear dilaton solution we considered
above, the coupling ultimately grows large in the very far past, making the
theory difficult to study.\foot{Note that this issue is
not special to the supercritical corners of string theory; the presence of a past
singularity at the level of the low energy effective theory pertains to all nontrivial cosmological backgrounds, including ones formulated in the critical theory.}
{\it A priori}, there are two natural approaches to this problem: (1) change
the set-up to maintain weak string
coupling throughout, and investigate possible resolutions of the singularity via
$\alpha'$ effects;  or (2) attempt to find a non-perturbative completion, perhaps
building on existing frameworks such as \refs{\bfss,\AdSCFT}.

\subsubsec{Weak string coupling at early times?}

Regarding approach (1), one might consider taking the early-time theory as a linear dilaton solution with
$g_s\sim e^{+\sqrt{2c_{eff}/3}X^0}$ {\it growing} in time, combined with the rolling tachyon solution. The
tachyon generically grows slowly toward the past and rapidly toward the future. In this case, the string
coupling can be tuned arbitrarily weak during the epoch when the tachyon is small, while it is not sourced when
the tachyon is large since the theory becomes a critical string compactification. However, this scenario is
subject to further instabilities -- the {\it pseudotachyonic} modes \refs{\newD,\Hell}\ may cause significant
backreaction in the solution with the growing string coupling if they have an infinite time to develop.\foot{We
thank O. Aharony and E. Witten for discussions of this case (as well as other aspects of the duality).} Their
effects are worth studying further.

%
%A priori there are two approaches to this problem.  One may (1) change the setup to maintain perturbativity
%throughout, investigating the possibility of $\alpha'$ effects resolving the singularity,  or (2) attempt to
%find a non-perturbative completion, perhaps building on existing frameworks such as \refs{\bfss,\AdSCFT}.
%Regarding possibility (1), one option that deserves further study is the linear dilaton solution with $g_s\sim
%e^{+\sqrt{2c_{eff}/3}X^0}$ (growing in time rather than shrinking) combined with the rolling tachyon solution.
%The tachyon generically grows slowly toward the past and rapidly toward the future. In this case, time evolution
%and RG flow are very different in the far future, but the string coupling can be tuned arbitrarily weak during
%the epoch when the tachyon is small (and it is not sourced when the tachyon is large). This however is subject
%to further instabilities -- the {\it pseudotachyonic} modes \refs{\newD,\Hell}\ which appear innocuously in the
%solution with $g_s\sim e^{-\sqrt{2c_{eff}/3}X^0}$ (as their effects die off with the decreasing string coupling)
%may cause significant back reaction in the solution with the growing string coupling if they have an infinite
%time to develop.  Their effects are worth studying further\foot{We thank O. Aharony and E. Witten for
%discussions of this case (as well as other aspects of the duality).}; in any case there may be variants of this
%idea, perhaps involving effects like \TE, which could yield an $\alpha'$ resolution of the initial singularity
%in these theories.

\subsubsec{A gauge theory dual}

An avenue for approach (2) is the AdS/CFT correspondence, which has provided a useful nonperturbative
formulation of a large class of backgrounds of superstring theory in the critical dimension\foot{This includes
recent proposals for resolving spacetime singularities using AdS/CFT \refs{\sing,\gary}\ (see also \CrapsWD\ in
the context of \bfss).}. To date, no such formulation has been derived for supercritical strings. We propose
such a formulation here. The general strategy is to begin with a known AdS/CFT dual and compactify the field
theory side on a time-dependent background with compact, negatively curved spatial slices, such that the
spacetime dual will also contain compact, negatively curved spatial slices. We have seen here and in
\refs{\mutation,\newdims}\ that superstring theory with such factors in spacetime is supercritical.

Consider the Poincar\'e patch of $AdS_5\times S^5$, dual to $U(N)$ ${\cal N}=4$ on Minkowski space ${\cal
M}_{3,1}$, and consider the foliation of ${\cal M}_{1,3}$ by hyperbolic 3-space $\IH_3$. Next, take an orbifold
of both sides of the duality by a discrete isometry group $\Gamma$ of $\IH_3$.  This has compact spatial slices
$\M_3=\IH_3/\Gamma$ as the fundamental domain of $\Gamma$, with a fundamental group of exponential growth whose
abelianization is the first homology group. The CFT will then propagate on the spacetime
\eqn\minkorb{
    ds^2=-dt^2+t^2ds^2_{\IH/\Gamma}\ ,
}
with the constant curvature metric descending from the metric $ds^2_{\IH}=dy^2+\sinh^2yd\Omega^2$ on the
covering space $\IH$. The string theory dual is type IIB on a patch of $AdS_5\times S^5$ covered by
\eqn\gravmet{ds^2={r^2\over
\ell^2}\bigl(-dt^2+t^2ds^2_{\IH/\Gamma}\bigr)+\ell^2 {dr^2\over r^2}
+ds^2_{S^5}}
with the spacetime curvature supported by the standard
$N$ units of
self-dual five-form Ramond-Ramond
(RR) flux.
% $F_5=N\epsilon_{\mu_1\dots\mu_5}$.
%The AdS radius $\ell$ is related
%to the number of branes $N$ and string coupling
% $g$ via $\ell^4 = (4\pi gN)l_s^4$.

In the spacetime theory, the volume ${\cal V}_3$ of $\M_3$ varies both
with time and with the radial direction. At fixed time $t$, ${\cal V}_3$
grows as $r\to\infty$, towards the boundary of our $AdS$ orbifold.
This is dual to the the UV regime of the CFT, in which the modes are not
sensitive to the compactness of the spatial slices.  ${\cal V}_3$ shrinks as
$r\to 0$; this is dual to the infrared regime of the CFT.  The discussions here and
in \refs{\mutation,\newdims}\ indicate that the effective central charge
of the spacetime theory grows larger as ${\cal V}_3$
shrinks.  Thus, the {\it infrared} physics of the CFT characterizes the
most {\it supercritical} regime of the string theory dual.

At this stage there are still barriers to realizing the supercritical theory more concretely. The field theory
lives on a compact space with volume ${\cal V}_3 = L(t)^3 = t^3$.  At energy scales of order the Hubble
expansion scale $1/t$, we expect the field theory to develop a mass gap and spacetime to end. Furthermore, the
computations in
\newdims\ and in \S2-3 were valid in perturbative string theory in the absence of RR flux. These problems can be
averted by taking the additional step of studying the theory in a state
%going out
on the approximate Coulomb branch of the CFT. We consider the near-horizon limit of $N$ $D3$-branes, placed on
an $SO(6)$-invariant shell at some radial position $R$, in the directions transverse to the worldvolume. The
branes source all $N$ units of 5-form flux, so inside the shell the flux is absent.\foot{Configurations like
this have been discussed in, \eg, \refs{\KrausHV, \GiddingsZU} (see also \DanielssonFA\ for a different type of
shell) and recently applied to a Scherk-Schwarz compactification of the CFT in \gary, whose conventions we
follow.} The static metric is $AdS_5\times S^5$ outside the shell, and ten dimensional flat space inside. After
the orbifold, the spacetime metric is:
\eqn\shell{ ds^2 =
h^{-1}(r)[-dt^2  + t^2ds^2_{\IH/\Gamma}] + h(r)[dr^2 + r^2
d\Omega_5^2]}
where $i=1,2,3$, and where
\eqn\Hshell{ h(r) = {\ell^2\over r^2} \quad (r>R), \qquad h(r) = {\ell^2\over R^2} \quad (r<R)\ .}
Thus, after the orbifold, the spacetime at small $r$ is preserved.

 Inside the shell, for all $r<R$, is a regime with no RR flux and with an
 $\M_3$ factor, of proper size $tR/\ell$, expanding with time.  The results
 of \S2-3\ and of \refs{\mutation,\newdims}\ indicate that this region is supercritical.
 According to the AdS/CFT dictionary, its physics is dual to the low-energy physics of the CFT on the spacetime \minkorb, on the Coulomb branch.

 If $\dim H_1(\M_3) = b_1$, the results of \S2\ show that the minimal
 completion of string theory on $\M_3$ at small $r$ or $t$ is a ``D-dual"
 torus $T^{b_1}$.  The T-dual of this torus appears naturally in the gauge theory
 as follows.
 Since all $N$ D-branes lie on the shell, the charged $W$-bosons are massive
 and low-energy dynamics of the theory is that of the $U(1)$ factors.  More precisely, there is a distribution
 of masses of $W$ bosons determined by the relative positions of the branes on the shell.  The number $n$ of
 D-branes close enough to a given brane to yield masses $m_W<1/L$ is given by
\eqn\nscale{n\sim N {l_s^{10}\over{(RL)^5}}}
We consider $RL\gg l_s^2$ so that this number is parameterically smaller than $N$.

 In the regime of small $g_sn$
%and large higgs vev (large $R$ in the above solution),
% We claim that by setting $g_s$ to be sufficiently small and $R$ to be sufficiently
% large,
we find that the low-energy dynamics includes Wilson lines in a range of scales between $\sqrt{g_sn}/L\ll E\ll
1/L$.
 These Wilson lines take values in
 ${\rm Sym}^N\left(T^{b_1} \right) \equiv (T^{b_1})^N/S_N$, with a $T^{b_1}$ factor for each $U(1)$.

To see that the Wilson lines are relevant modes at low energies,
one must first check that higher momentum modes
on $\M_3$
 do not contribute at low energies.  The Wilson lines live on a compact space, and yield quantum mechanical
 energy levels $E_j\sim j^2g_s/L$.
%Because the kinetic term has a factor of
% $1/g^2_{YM} \sim 1/g_s$, the kinetic energy
% of the Wilson lines is of order $g_s/L$.
 For sufficiently weak string coupling,
 this is much smaller than the scale $1/L$ of higher momentum modes of the gauge theory.

 Secondly,
the moduli space of Wilson lines is approximate; massive $W$
 bosons contribute a potential, as studied for example in \axionpot.
At low energies, we must check that this potential is small compared to the scale $1/L$. We can ensure this by
making $R$ large so that the number $n$ of light
 $W$ bosons
 charged under a given $U(1)$ \nscale\ is small, and most $W$ bosons
 are heavy compared to the scale $1/L$
 of momentum modes on $\M_3$.  The resulting potential energy scales like $g_sn\phi^a/L^2$ where $\phi_a$ is
 the canonically normalized Wilson line scalar: that is, the mass$^2$ of the Wilson lines is much
 smaller than the KK scale $1/L$.  This same effect suppresses the potential on the Coulomb branch
 itself, and the dynamics of the scalars in the $4d$ gauge theory should also be taken into account.

  We conclude that at low energies of the gauge theory, the gauge dynamics
 includes the motion of $N$ $U(1)$ Wilson lines on an approximate moduli space
${\rm Sym}^N\left(T^{b_1} \right) $.
 The negative curvature
 of the spatial slices provides Hubble friction slowing down motion on the
 approximate moduli space. Note that the volume of this $T^{b_1}$ moduli
 space {\it grows}\ with decreasing ${\cal V}_3$.
 %We believe that
The torus described by
the Wilson lines is
 the T-dual of the minimal completion derived in \S2\ from current
 algebra considerations. It may be tempting to
 identify the inevitable potential on the moduli space, arising from integrating out
 the $W$-bosons, with the tachyon potential described in \S2-3; however,
 under T-duality it maps to a condensate of winding modes.  The appearance of the
 tachyon potential of \S2-3\ from the gauge theory is an interesting question that
 we leave for future work.

 We emphasize again that in this set-up, the $T^{b_1}$ arises in the {\it infrared} regime
 of the CFT, where considerations of universality and naturalness apply;
 whereas it appears in the {\it ultraviolet} regime of the worldsheet matter
 sigma model.  This suggests that the minimal completion of the system
 preserving the winding symmetries indeed arises naturally, in the Wilsonian sense,
 from the gauge theory dual.  At least it implies that the other trajectories of the worldsheet RG
 which miss the UV fixed point of the minimal completion
 are themselves closely related to a theory on $T^{b_1}$, as argued in \S3.1.
 The range of possible trajectories involving the $T^{b_1}$
 may be mirrored in the dual gauge theory by the
 ambiguity in the choice of state on the initial lightcone of singularities
 at $t=0$ in \minkorb.

 The conformal transformation of \minkorb\ to
 \eqn\conftrans{-d\tau^2+ds^2_{\IH/\Gamma}}
(where $\tau=log(t/t_0)$)
 may help elucidate the early-time physics, though its effect on the scalar field dynamics must also be
 included consistently.\foot{We thank D. Berenstein and E. Witten for this suggestion.}  In the conformal
 frame \conftrans, the conformal
 coupling $R\phi^2$ contributes a negative effective mass squared for the scalars in the field theory \refs{\eg\ \WittenXP}.
 Indeed, in the conformally rescaled background, $\phi\propto t\propto e^{\tau/\tau_0}$ grows
 like a tachyon.  However, in the original conformal frame \minkorb,
 the curvature is zero so the conformal coupling $R\phi^2$
vanishes. The late time behavior is simply that of
 a quantum field theory on a slowly expanding FRW background with compact hyperbolic spatial slices.  Thus
 it appears that the instability of the CFT on \conftrans\ leads to an innocuous endpoint:  field
 theory on a late-time weakly curved cosmological spacetime.  Conversely, the singularity in \minkorb\ as
 one goes back to $t=0$ translates into the return of $\phi$ to the origin of the moduli space
 as $\tau\to -\infty$ in the frame \conftrans.
 It will be interesting to explore the consequences of this formulation of the
 system via AdS/CFT.

\vskip .5cm

\noindent{\bf Acknowledgements}

\nobreak We are grateful to S. Hellerman and M. Ro\v{c}ek for detailed discussions of the
worldsheet-supersymmetric case. We thank O. Aharony, D. Berenstein, R. Dijkgraaf, M. Gaberdiel, S. Kachru, D.
Long, M. Matone, H. Schnitzer, A. Sever, S. Shenker,  D. Starr, P. Storm,  D. Tong, C. Vafa, E. Verlinde, and E.
Witten for very useful discussions. D.G., A.L., D.R.M., and E.S. thank the KITP for hospitality during part of
this work. This research was supported in part by the National Science Foundation under Grant No. PHY99-07949.
Any opinions, findings, and conclusions or recommendations expressed in  this material are those of the authors
and do not necessarily reflect the views of the National Science Foundation. D.G. is supported in part by an
NSERC fellowship. D.G. and E.S. are also supported in part by the DOE under contract DE-AC03-76SF00515, by the
NSF under contract 9870115, and by an FQXI grant. The work of A.L. is supported in part by the NSF grant
PHY-0331516, by the DOE grant DE-FG02-92ER40706, and by an Outstanding Junior Investigator award.  The work of
J.M. is supported in part by funds provided by the U.S. Department of Energy (D.O.E.) under cooperative research
agreement DE-FG0205ER41360. The work of D.R.M. is supported in part by National Science Foundation grant
DMS-0606578.

\listrefs

\end